\documentclass[aps,prb,superscriptaddress,twocolumn,showpacs]{revtex4-2}
\usepackage{amssymb}
\usepackage{epsfig}
\DeclareMathAlphabet{\pazocal}{OMS}{zplm}{m}{n}            
\usepackage{graphicx}
\usepackage[usenames, dvipsnames]{xcolor}
\usepackage{bm}
\usepackage[normalem]{ulem}     
\usepackage{soul}
\usepackage{amsmath}
\usepackage{braket} 
\usepackage{rotating}
\usepackage{multirow} 
\usepackage{mhchem}
\DeclareMathAlphabet{\pazocal}{OMS}{zplm}{m}{n}            
\usepackage{graphicx}
\usepackage{hyperref}
\usepackage[normalem]{ulem}
\usepackage{float}

\usepackage{tikz,xcolor,hyperref}
\usepackage[normalem]{ulem}

\definecolor{lime}{HTML}{A6CE39}
\DeclareRobustCommand{\orcidicon}{%
	\begin{tikzpicture}
	\draw[lime, fill=lime] (0,0)
	circle [radius=0.16]
	node[white] {{\fontfamily{qag}\selectfont \tiny ID}};
	\draw[white, fill=white] (-0.0625,0.095)
	circle [radius=0.007];
	\end{tikzpicture}
	\hspace{-2mm}
}

\foreach \x in {A, ..., Z}{%
	\expandafter\xdef\csname orcid\x\endcsname{\noexpand\href{https://orcid.org/\csname orcidauthor\x\endcsname}{\noexpand\orcidicon}}
}

\newcommand\redsout{\bgroup\markoverwith{\textcolor{red}{\rule[0.5ex]{2pt}{0.4pt}}}\ULon}
\newcommand\bluesout{\bgroup\markoverwith{\textcolor{blue}{\rule[0.5ex]{2pt}{0.4pt}}}\ULon}

\newcommand{\SPhide}[1]{{}}

\begin{document}
\title{Bulk photovoltaic effect in MoSe$_2$ and  Janus MoSSe sliding ferroelectrics}     

\author{Roumita Roy \orcidA}
\email{roumita95@gmail.com}
\affiliation{Department of Materials Science, University Milan-Bicocca, 20125 Milan, Italy}

\author{Giuseppe Cuono\orcidB}
\affiliation{Consiglio Nazionale delle Ricerche (CNR-SPIN), Unit\'a di Ricerca presso Terzi c/o Universit\'a “G. D’Annunzio”, 66100 Chieti, Italy}

\author{Silvia Picozzi\orcidC}
\affiliation{Department of Materials Science, University Milan-Bicocca, 20125 Milan, Italy}
\affiliation{Consiglio Nazionale delle Ricerche (CNR-SPIN), Unit\'a di Ricerca presso Terzi c/o Universit\'a “G. D’Annunzio”, 66100 Chieti, Italy}

\date{\today}

\begin{abstract}
We present a first-principles study of the nonlinear optical properties of sliding ferroelectric bilayers based on MoSe$_2$ and Janus MoSSe. Two Janus configurations are considered: {\em i)} one bilayer where the two intralayer polarizations caused by Janus chemical asymmetry cancel each other out, yielding photocurrent spectra comparable to pristine MoSe$_2$ bilayers; {\em ii)} another bilayer where the intralayer polarizations add up, for which the photoresponses are strongly enhanced. Our results show that photocurrent generation in the polar Janus structures is predominantly governed by vertical chemical asymmetry, with limited dependence on the sliding direction. These findings highlight complementary design strategies: interlayer sliding enables sensitivity to external  tuning, while the Janus  intralayer polarization enhances photoresponses in the visible range. The interplay between composition and stacking therefore provides a versatile platform for tailoring light–matter interactions in 2D ferroelectric materials.
\end{abstract}

\maketitle

\textit{Introduction-} The efficient conversion of solar energy into electrical power remains a central challenge in the development of next-generation optoelectronic and energy-harvesting technologies. Conventional photovoltaic devices rely on p–n junctions, where built-in electric fields separate the light induced charge carriers. In contrast, the bulk photovoltaic effect (BPVE) is observed in materials that lack inversion symmetry \cite{Dai,Fridkin2000,Glass1974,vonbaltz1981,Sipe}. The main advantage of such systems is that they offer a mechanism for photocurrent generation in the absence of external bias or heterojunctions, which simplifies large scale device fabrication. Although several studies have reported that the BPVE can produce open-circuit voltages above the material band gap by enabling carrier collection prior to thermalization, which could surpass the Shockley-Queisser limit \cite{Spanier2016}, others have questioned this claim, citing additional factors that may restrict the overall energy conversion efficiency \cite{Kirk2017,Pusch2023}. Despite these opposing perspectives, BPVE continues to stimulate significant theoretical \cite{Young2012,SC,IC,Cook2017,Zhang2019,Zhang2018} and experimental \cite{Cote2002,Nakamura2017,Rees2020,McIver2012,Yuan2014, Niu,Lin2024,Urakami,Song2025} interest due to its intrinsic relation to band structure and its ability to generate ultrafast, bias-free photocurrents \cite{Dai,Ahn2020,Morimoto2016,Parker2019,He2024,Rangel2017,QMa}. 

BPVE arises from second-order nonlinear optical processes and is composed of two key mechanisms: the shift and the injection current \cite{Sipe,Dai,Sturman}. The shift current originates from the position shifts of charge centers during interband transitions \cite{Dai,vonbaltz1981,Young2012,nastos2010,resta2024,SC}, while the injection current arises from asymmetries in the group velocity distribution of excited carriers \cite{Dai,Zhang2019,Dai2021,IC,Stavric2025}. These mechanisms are strongly influenced by the crystal symmetry and electronic structure of the material. For non-magnetic materials, the bulk photovoltaic response consists of linear shift current and circular injection current. In contrast, linear injection current requires broken time-reversal symmetry and thus arises only in magnetic systems. 

\begin{figure*}
\begin{center}
1\includegraphics [width= \textwidth] {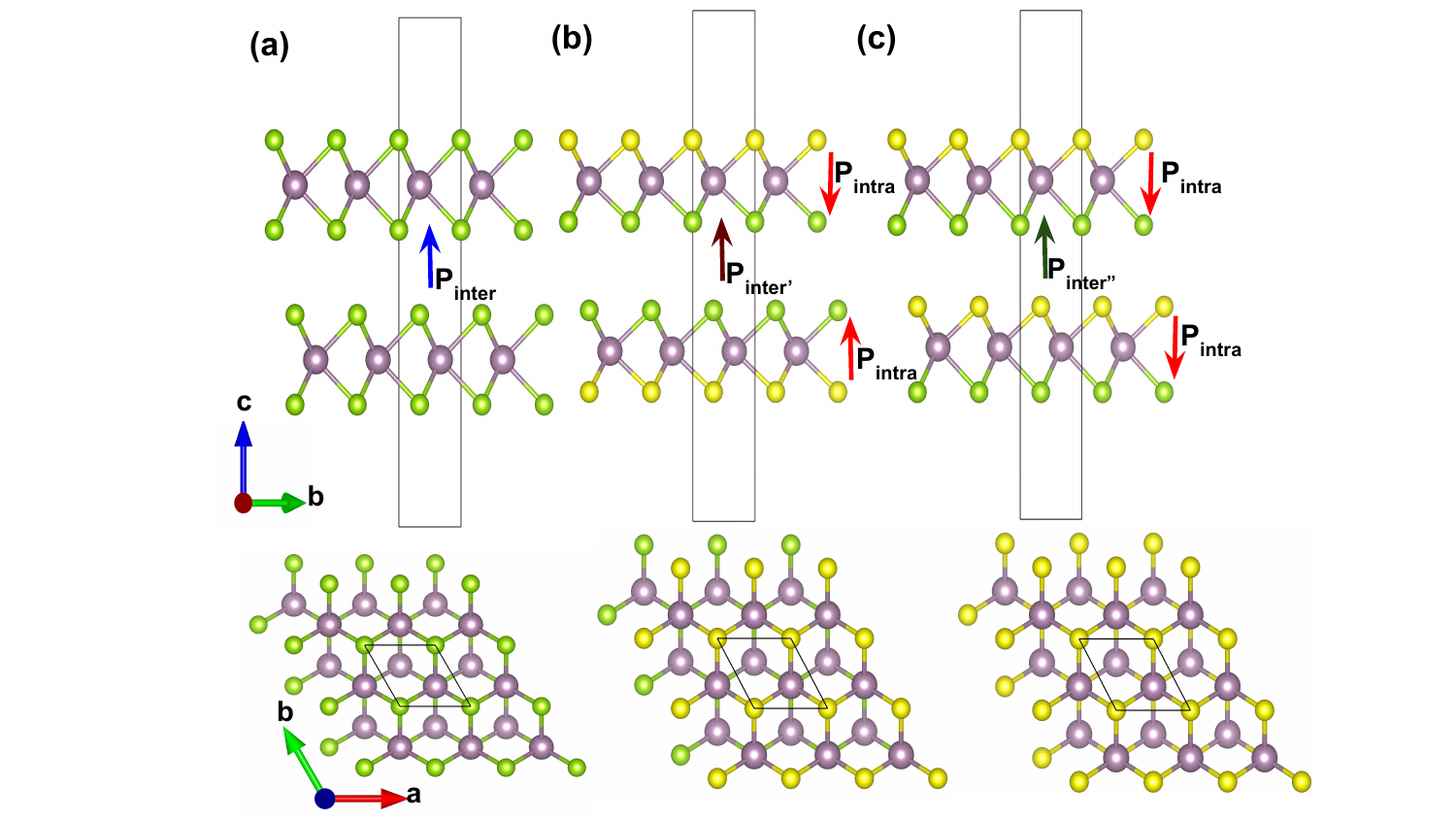}
\end{center}
\caption{ The side and top views of the bilayer crystal structures of (a) MoSe$_2$, (b) Janus I, and (c) Janus II, respectively. The Mo, Se and S atoms are represented by violet, green,  and yellow spheres respectively. The blue, maroon, and green arrows represent the direction of the interlayer polarization (P$_{inter}$/P$_{inter'}$/P$_{inter"}$) for MoSe$_2$, Janus I, and Janus II respectively. The red arrows represent the direction of intralayer polarization (P$_{intra}$) exisisting within the Janus I and Janus II monolayers.}
\label{fig:Fig1}
\end{figure*} 

Due to their inherent broken inversion symmetry and spontaneous, switchable polarization, ferroelectric materials have been considered prototype systems for studying the BPVE \cite{Young2012,Young,Butler,Tan2016,Pandey2019,Tiwari,SBS,Ogawa2017,Nakamura2018,SbSI,Nakamura2017}. The photoresponse typically switches along with the ferroelectric polarization, making the electric field a practical control knob for tuning the direction of the photocurrent \cite{Dai,Kim2019,Choi2009,Ji2010,Grinberg2013}.
It was also shown that the BPVE can be significantly enhanced in low-dimensional systems such as strain-engineered MoTe$_2$ \cite{MoTe2}, ReS$_2$ \cite{ReS2}, group-IV monochalcogenide monolayers \cite{Rangel2017,Panday} and other two-dimensional ferroelectrics \cite{Qian2023,Li2021}, where the enhancement is often attributed to van Hove singularities in the electronic density of states.

Recently, the discussion of BPVE was extended to a class of materials known as sliding ferroelectrics \cite{Zhang2025,Fei2018,BN21,Barrera2021,Stern2021,Hu2019,Hu2021,Wang2022,Weston2022,Meng2022,Sui2023,Wang2023,Sun2025,LinNano}.
In these systems, the polarization arises from interlayer charge redistribution in stacking arrangements lacking inversion symmetry. It was shown that even if the monolayers are centrosymmetric, it is possible to build bilayers which break inversion symmetry and generate out of plane polarization \cite{Zhang2025}. 
In sliding ferroelectrics, the polarization is fully determined by the interlayer stacking arrangement and can be tuned by diverse external stimuli, offering great potential for advanced information storage and processing applications.
Furthermore, the interlayer sliding mechanism enables reversible polarization switching with potentially lower energy barriers and minimal structural distortion \cite{Chen2024}. From a BPVE perspective, such tunability allows dynamic modulation of the nonlinear optical response and may enhance shift and injection currents through engineered stacking configurations, strain control or magnetic order \cite{Xiao2022,Zhang2022,Ji2025,Liang2025}. 
In 2D sliding ferroelectrics, a non-synchronous BPVE coupled to the ferroelectric order has been identified \cite{Xiao2022}: while the in-plane BPVE components, which correspond to photocurrent induced in the xy plane, are invariant under polarization reversal, the out-of-plane components, corresponding to photocurrent induced along the z direction, switch sign \cite{Xiao2022}. The unswitchable in-plane responses could be advantageous for large-scale photoelectric applications. 

In this paper, we perform first-principles calculations based on density functional theory (DFT) to investigate the BPVE in sliding ferroelectric bilayers of MoSe$_2$ and Janus MoSSe. Janus monolayers inherently lack mirror symmetry due to their vertical compositional asymmetry, which leads to out-of-plane dipole moments and enhanced piezoelectric  responses \cite{Peeters,Pham,Zhang2024}.
Investigating BPVE in Janus bilayers could help identify possible strategies to enhance nonlinear photovoltaic effects in two-dimensional systems. We study two Janus bilayer configurations: one where the intralayer polarizations induced by Janus asymmetry cancel each other, and another exhibiting a net intralayer polarization. In the latter case, we find that the photocurrents are significantly enhanced compared both to the former and to the MoSe$_2$ bilayer. While sliding ferroelectricity allows for easy switching of photoresponses, this polar Janus structure is particularly promising for applications, as it generates very large photocurrents exceeding those reported for many other 2D and 3D materials.

\begin{figure*}
\begin{center}
\includegraphics [width= 16cm] {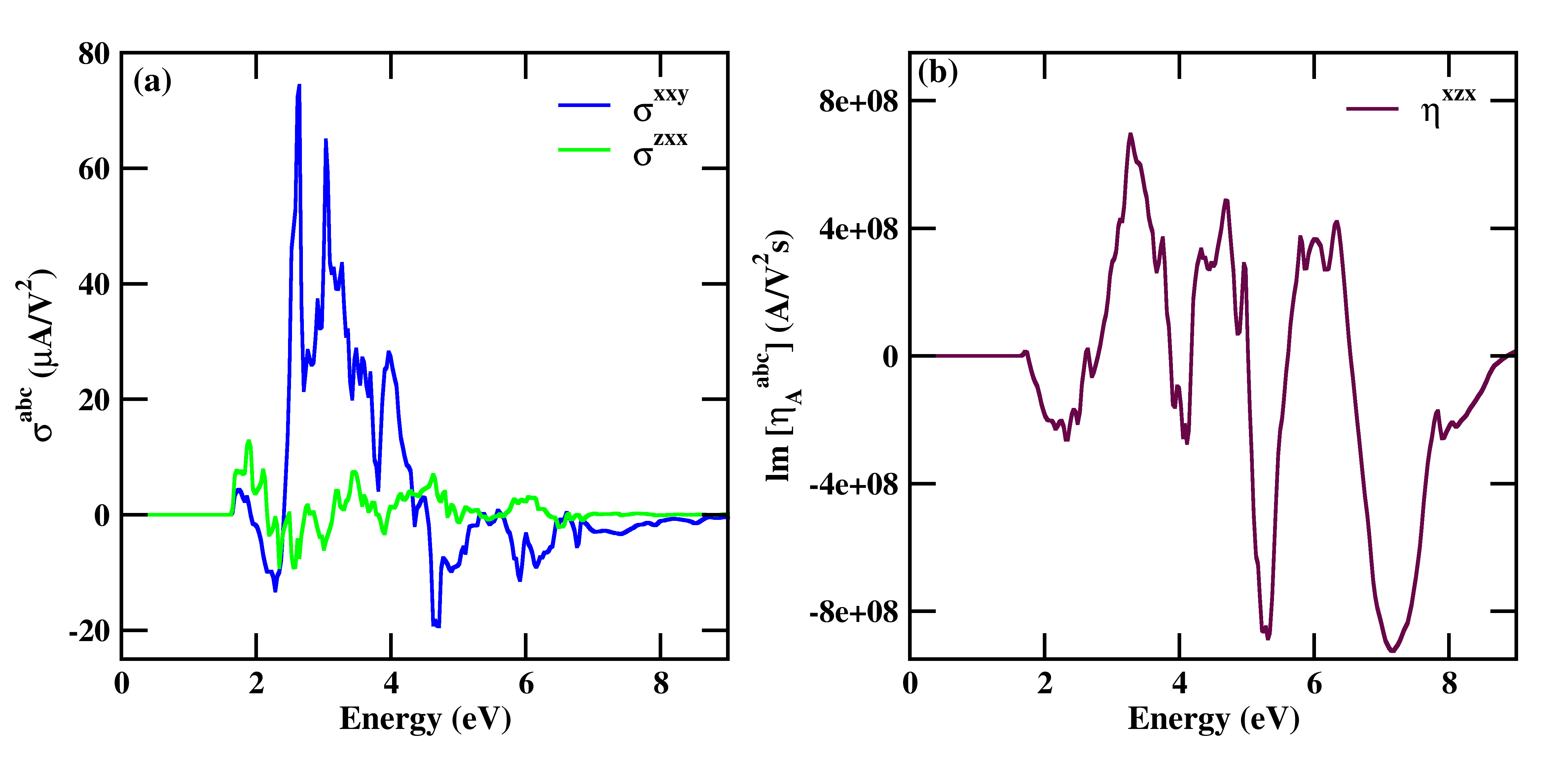}
\end{center}
\caption{The non-zero, independent components of the photoconductivity tensor for MoSe$_2$. (a) Shift current components (b) Circular injection current components.}
\label{fig:Fig3}
\end{figure*} 

\textit{Computational Details-} We performed first-principles calculations using DFT within the projector augmented-wave (PAW) framework, as implemented in the VASP code \cite{Kresse1993,Kresse,Kresse1996}. The exchange-correlation energy was treated using the generalized gradient approximation (GGA) in the Perdew–Burke–Ernzerhof (PBE) form \cite{Perdew1996}. Spin–orbit coupling (SOC) was included in all the simulations, except during structural optimization. A plane-wave kinetic energy cutoff of 400 eV was used throughout and we employed a $20\times20\times1$ Monkhorst-Pack $k$-grid centered
at the ${\Gamma}$ point. To prevent artificial interactions between periodic images, a vacuum spacing of $\approx$ 35 {\AA} was introduced in our structures. The atomic positions were optimized until the residual forces were smaller 0.001 eV/{\AA}. To accurately capture interlayer interactions inherent to van der Waals layered systems, dispersion corrections were included via the DFT-D3 scheme of Grimme \cite{Vdw}. 

To compute the bulk photovoltaic response, including both shift and injection current contributions, we followed the method developed by Ibañez-Azpiroz $et$ $al.$ \cite{SC} and by Puente-Uriona $et$ $al.$ \cite{IC}, which rely on Wannier interpolation of the DFT electronic structure.  The maximally localized Wannier functions (MLWFs) \cite{W901,W902} were generated using the WANNIER90 package \cite{W903}. The DFT total band structure was projected on the Mo-$d$ and chalcogen (S,Se)-$p$ orbitals to accurately capture low energy physics. A very dense $k$-point grid up to $500\times500\times1$ was employed for the optical calculations to ensure convergence of the computed photoconductivities. A fixed broadening of 0.02 eV was applied to effectively handle van Hove singularities, following the prescription of \cite{SC}. Following previous Refs. \cite{SC,Rangel2017,Panday,Tiwari,Xiao2022}, the renormalized 3D-like responses were obtained by assuming a single active layer with $\sigma_{\text{3D}} = \frac{c}{L_{\text{active}}} \, \sigma_{\text{slab}}$, where $c$ is the lattice parameter, $L_{active}$ is the thickness of the bilayer and $\sigma_{\text{slab}}$ is the calculated BPVE coefficient.

\textit{Crystal Structure-} The monolayers used to construct the MoSe$_2$ bilayers belong to the 3R polytype. They exhibit trigonal prismatic coordination around the transition metal atoms and are noncentrosymmetric, nonpolar, and possess horizontal mirror symmetry \cite{Xiao2022}.

To construct the sliding ferroelectric bilayer, two noncentrosymmetric MoSe$_2$ monolayers are vertically stacked in an AA-type configuration, where each layer is identical and A denotes the monolayer structure, as shown in Fig. \ref{fig:Fig1}(a), with the top and bottom layers laterally slid relative to each other.  The bilayer belongs to the space group P3m1 (No. 156) and this system exhibits a net out-of-plane polarization, signaling the emergence of ferroelectricity induced by interlayer sliding along the in-plane direction. The out-of-plane ferroelectric polarization can be reversed through in-plane interlayer displacement  and the two ferroelectric states with opposite polarization have a low switching barrier,  and are related by mirror symmetry \cite{Xiao2022}. 

Next, we construct Janus bilayers by replacing the chalcogen atoms in one Se sublayer of each monolayer with S. Two distinct bilayer configurations are examined, which we refer to as Janus I and Janus II, shown in Fig. \ref{fig:Fig1} (b) and (c), respectively. In the Janus I arrangement, both the bottom of the top monolayer and the top of the bottom monolayer are terminated with Se atoms, forming a S-Mo-Se–Se-Mo-S stacking.  In this configuration, the intralayer polarizations induced by the Janus asymmetry are oriented in opposite directions and effectively cancel each other out within the bilayer. In contrast, the Janus II structure introduces structural asymmetry with a Se-Mo-S–Se-Mo-S stacking, leading to an intrinsic intralayer out-of-plane polarization even in the absence of interlayer sliding.
Substituting S for Se also induces local structural distortions arising from differences in atomic radius and electronegativity: the Mo–Se bond length in pristine MoSe$_2$ bilayers (2.51 {\AA}) changes to 2.40 {\AA} for Mo–S and 2.52 {\AA} for Mo–Se in Janus I. The distortion is more pronounced in Janus II, where the Mo–S bonds measure 2.61 {\AA} and 2.40 {\AA} for the top and bottom monolayers respectively, further modifying the internal electric field across the sliding bilayers. As for the evaluation of polarization, we calculated the polarization for MoSe$_2$ and found a value which agrees well with the previously reported result in \cite{Zhang2024}, namely $\approx$ 0.7 pC/m. The polarization magnitude in Janus I is of the same order as that of MoSe$_2$, being $\approx$ 0.5 pC/m, whereas in Janus II it is two orders of magnitude larger, namely $\approx$ 20.0 pC/m, dominated by the intralayer contribution \cite{Zhang2024}.

\textit{BPVE in sliding MoSe$_2$ bilayer-} To investigate the nonlinear optical response of the sliding ferroelectric MoSe$_2$ bilayer, we compute the second-order photoconductivity tensor elements that govern the BPVE, including both shift and injection current contributions. Regarding the shift current response $\sigma^{abc}$, the non-zero components allowed for the space group P3m1 can be categorized into in-plane and out-of-plane contributions, where the in-plane components correspond to photocurrents induced in the xy plane, while the out-of-plane components correspond to photocurrents induced along the z direction. 
 For the in-plane response, the component is given by $\sigma^{xxy}$ = $\sigma^{yxx}$ = -$\sigma^{yyy}$, whereas for the out of plane one we have $\sigma^{zxx}$ = $\sigma^{zyy}$. Other allowed components, namely $\sigma^{xxz} = \sigma^{yyz}$ and $\sigma^{zzz}$, are not shown because their contributions are negligible compared with the dominant ones. 

The $\sigma^{xxy}$ contribution is shown in Fig. \ref{fig:Fig3}(a), it exhibits prominent peaks at approximately 2.5 and 3 eV, with a maximum intensity of around  70 $\mu$A/V$^2$.
The out-of-plane shift current component \( \sigma^{zxx} \), also shown in Fig.~\ref{fig:Fig3}(a), remains large, even if smaller compared to the in-plane components, and features prominent peaks at around 1.9 eV and 2.5 eV, with largest intensity of about 15 $\mu$A/V$^2$. The values of both $\sigma^{xxy}$ and  \( \sigma^{zxx} \) in the visible range are comparable or larger than those of other ferroelectric materials \cite{SbSI,Young2012}, although smaller than the responses observed in various low-dimensional systems \cite{Rangel2017,Tiwari}.

For the circular injection current response $\eta^{abc}$, the allowed elements are $\eta^{xzx}$ = -$\eta^{yyz}$, with just one independent component.   
In Fig. \ref{fig:Fig3}(b), we show the circular injection current component $\eta^{xzx}$. The spectral response shows intense peaks of about 3 $\times$ 10$^8$ A/(V$^2$s) and 7 $\times$ 10$^8$ A/(V$^2$s) in the visible region, as well as pronounced peaks at larger energies. These values are larger than those reported for the Weyl semimetal TaIrTe$_4$ \cite{IC}, which exhibits a giant BPVE. Further, the in-plane components exhibit identical spectra for +P and –P states, whereas the out-of-plane components reverse sign upon polarization switching (see Supplemental Material \cite{Supplemental}). This behavior agrees with the symmetry analysis of Ref.~\cite{Xiao2022}, where mirror-related ferroelectric states lead to a non-synchronous BPVE by preserving specific tensor components.

\begin{figure}
\begin{center}
\includegraphics [width= 9 cm] {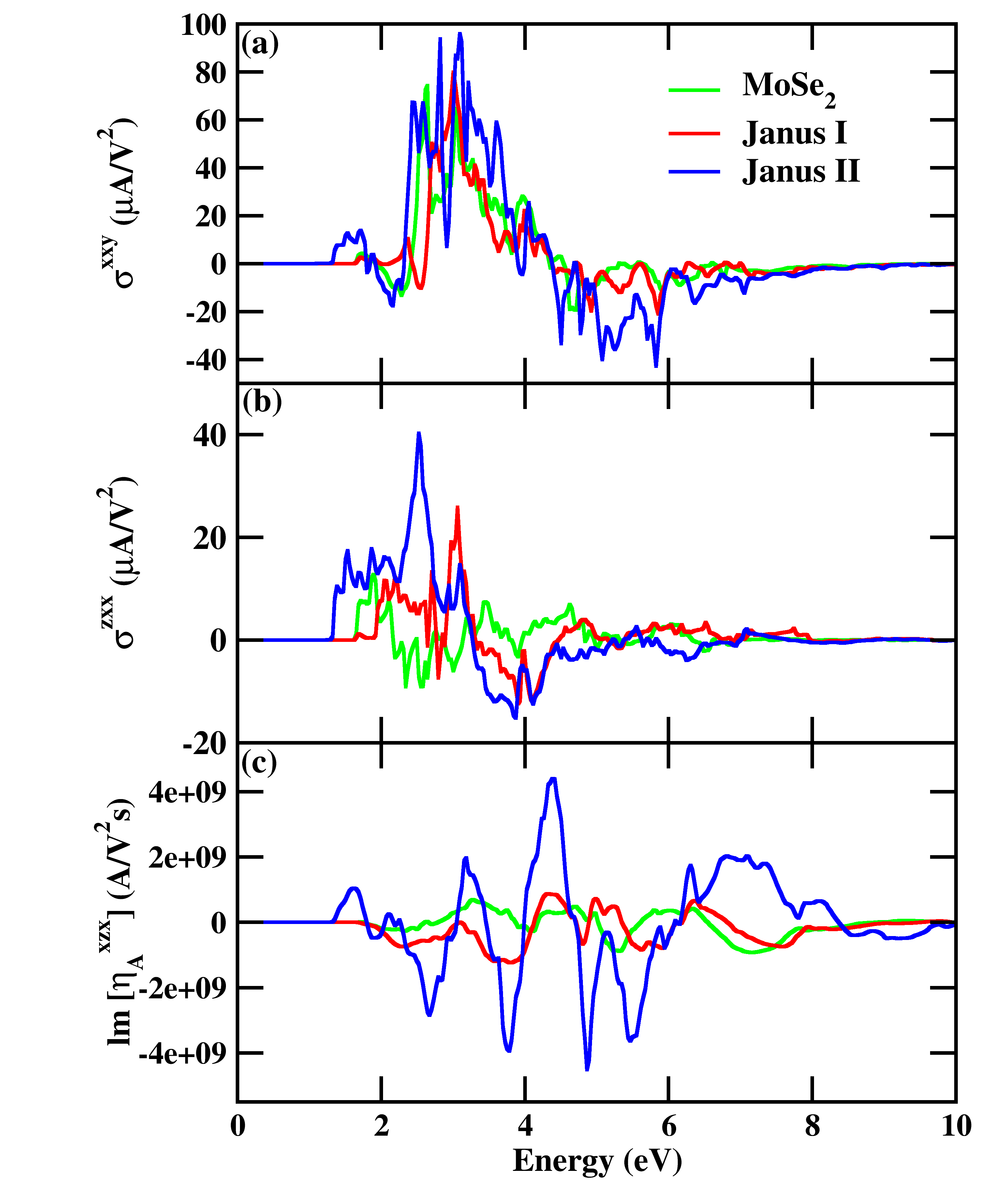}
\end{center}
\caption{Comparision of BPVE response for MoSe$_2$, Janus I and Janus II bilayers for a) in-plane shift current component $\sigma_{xxy}$ (b) out-of-plane shift current component $\sigma_{zxx}$ and (c) circular injection current component $\eta_{xzx}$.}
\label{fig:Fig4}
\end{figure}

\textit{BPVE in sliding Janus bilayers-} To elucidate the impact of chemical asymmetry on the BPVE of bilayers, we compare the shift current and the circular injection current components for MoSe$_2$, Janus I, and Janus II structures, as shown in Fig. \ref{fig:Fig4}. All the three structures belong to the space group P3m1 (No. 156), therefore we compare the same allowed components of the photocurrents.

Among the three, Janus II displays remarkable enhancement in all the components of the photocurrent. The $\sigma^{xxy}$ response of Janus II shows a sharp, well-defined peak at around 3 eV of about 90 $\mu$A/V$^2$, and maintains elevated values over a broad photon energy range, as shown in Fig. \ref{fig:Fig4}(a). In contrast, Janus I and MoSe$_2$ exhibit less intense largest peaks of around 80 $\mu$A/V$^2$ and 70 $\mu$A/V$^2$, respectively. 

The shift current component $\sigma^{zxx}$ and the circular injection current component $\eta^{xzx}$ show even more pronounced enhancements in the Janus II system.
As we can see from Fig. \ref{fig:Fig4}(b), $\sigma^{zxx}$ of Janus II configuration reaches peaks of around 40 $\mu$A/V$^2$ in the visible region, while the peaks of Janus I and MoSe$_2$ bilayers are much smaller, being around 10-20 $\mu$A/V$^2$ in the visible range.
Regarding the circular injection current, Janus II attains peak values nearly an order of magnitude greater than both MoSe$_2$ and Janus I, as shown in Fig. \ref{fig:Fig4}(c). The largest peaks are of around 4 $\times$ 10$^9$ A/(V$^2$s), with a very intense spectrum in both the visible and the ultraviolet regions. These values are of the same order of magnitude of the circular responses reported for EuO \cite{Stavric2025}, which was previously regarded as exhibiting a colossal BPVE in both its linear and circular components. Moreover, they exceed the values reported for several materials, both 3D and 2D \cite{Panday}, although they remain lower than those found in certain group-IV monochalcogenide monolayers \cite{Panday}. This suggests highly efficient photoresponses in the polar Janus II structure, making it a potentially interesting candidate for applications.

\begin{figure}
\begin{center}
\includegraphics [width= 7.5 cm] {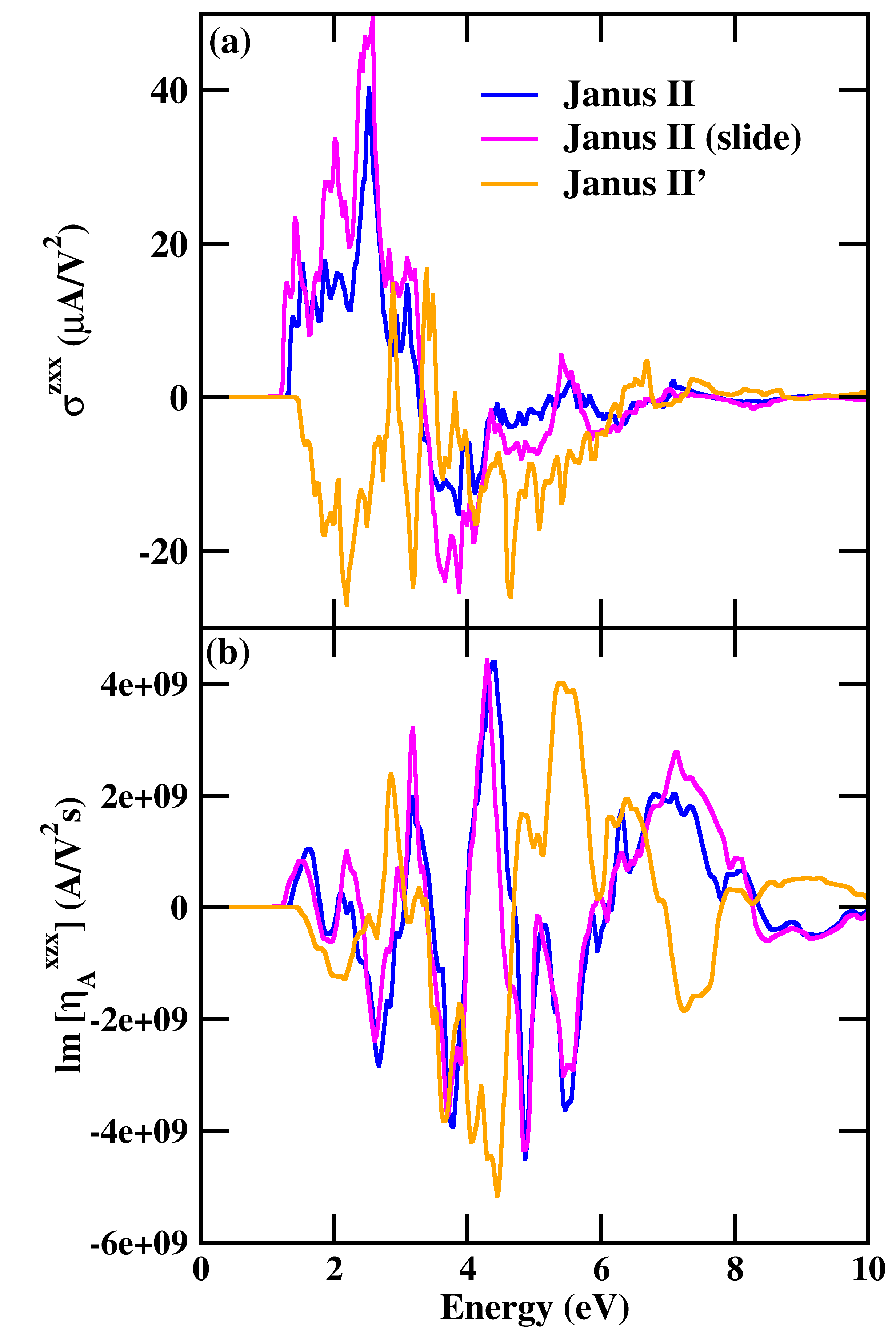}
\end{center}
\caption{Evolution of the (a) out-of-plane shift current $\sigma_{zxx}$ and (b) circular injection current  $\eta_{xzx}$ components for Janus II, Janus II (slide) and Janus II'.}
\label{fig:Fig5}
\end{figure} 

These findings underscore the strong tunability of the BPVE through chemical engineering. 
In Janus I, the opposing intralayer polarizations cancel each other out within the bilayer, resulting in similar band structure features (see Supplemental Material) and photocurrent intensities comparable to those of the sliding MoSe$_2$ bilayer. Janus II combines both intralayer polarization from chalcogen asymmetry and interlayer sliding ferroelectricity, leading to strongly enhanced bulk photovoltaic responses. To disentangle the relative importance of these two contributions, we analyze the evolution of photocurrents under controlled structural modifications, as illustrated in Fig. \ref{fig:Fig5}. First, we laterally slide the two monolayers relative to each other, thereby reversing the interlayer sliding polarization; we refer to this configuration as Janus II (slide). Second, we reverse the intralayer polarization by interchanging the S and Se atoms within each monolayer, creating a modified structure denoted as Janus II'. The structural details of Janus II (slide) and Janus II' can be found in the Supplemental Material.
Our results show that the photocurrent response is overwhelmingly dictated by the vertical chemical asymmetry of the Janus structure. Switching the intralayer polarization (Janus II') reverses both the shift and injection current components, underscoring the dominant role of the chemically driven dipole. In contrast, when only the sliding direction is modified, the photocurrent spectra exhibit only modest variations in intensity while retaining their original sign. These findings are consistent with recent reports on polarizations in these Janus systems \cite{Zhang2024}, which show that, when only the interlayer polarization switches, the overall polarization direction remains unchanged due to the dominant intralayer contribution.

\textit{Conclusions-} We investigated the BPVE in sliding ferroelectric MoSe$_2$ and Janus MoSSe bilayers. For the Janus structures, we considered two configurations: one where the intralayer polarizations induced by Janus asymmetry cancel each other, and another exhibiting a net intralayer polarization.
In the first case, photocurrent intensities are comparable to those of the MoSe$_2$ bilayer. In contrast, the Janus II shows significantly enhanced photoresponse, particularly in the circular injection current, which reaches values an order of magnitude higher than in the other cases. It exhibits pronounced photoresponses across the whole visible spectrum, making it a promising candidate for optoelectronic applications. Our analysis indicates that the BPVE in the polar Janus structure is predominantly driven by chemical asymmetry, with interlayer sliding polarization playing a smaller role.
As a result, in polar Janus structures, where intralayer mechanisms dominate, photocurrents cannot be simply inverted by reversing the sliding direction. While interlayer sliding ferroelectricity allows for easier switching of photocurrents, intralayer and chemically--induced polarizations enhance the photoresponses but they cannot be reversed.  These two mechanisms thus offer complementary functionalities: one enables tunable and reversible control, while the other provides strong enhancement of the response. 
Overall, chemical and structural engineering, particularly combining Janus-type asymmetry with ferroelectric stacking, provides a versatile strategy to design two-dimensional materials with highly responsive and tunable nonlinear optical properties.

\textit{Acknowledgments}
S.P. and R.R. acknowledge the financial support from the Next-Generation-EU program via the PRIN-2022 SORBET  (Grant No. 2022ZY8HJ). S.P. and G.C. acknowledge partial financial support by the PNRR Partenariato PE4 “National Quantum Science
and Technology Institute” (NQSTI) project PE0000023 and by the CANVAS project, funded by the Italian Ministry of the Environment and the Energy Security. HPC resources and support were provided by CINECA under the ISCRA IsB28 HEXTIM, IsCc2 SFERA, and IsCc9 BRIMS projects.

\clearpage
\onecolumngrid       
\clearpage
\section*{Supplemental Material}

\section{Electronic properties}
\begin{figure}[H]
\begin{center}
\includegraphics [width= 16 cm] {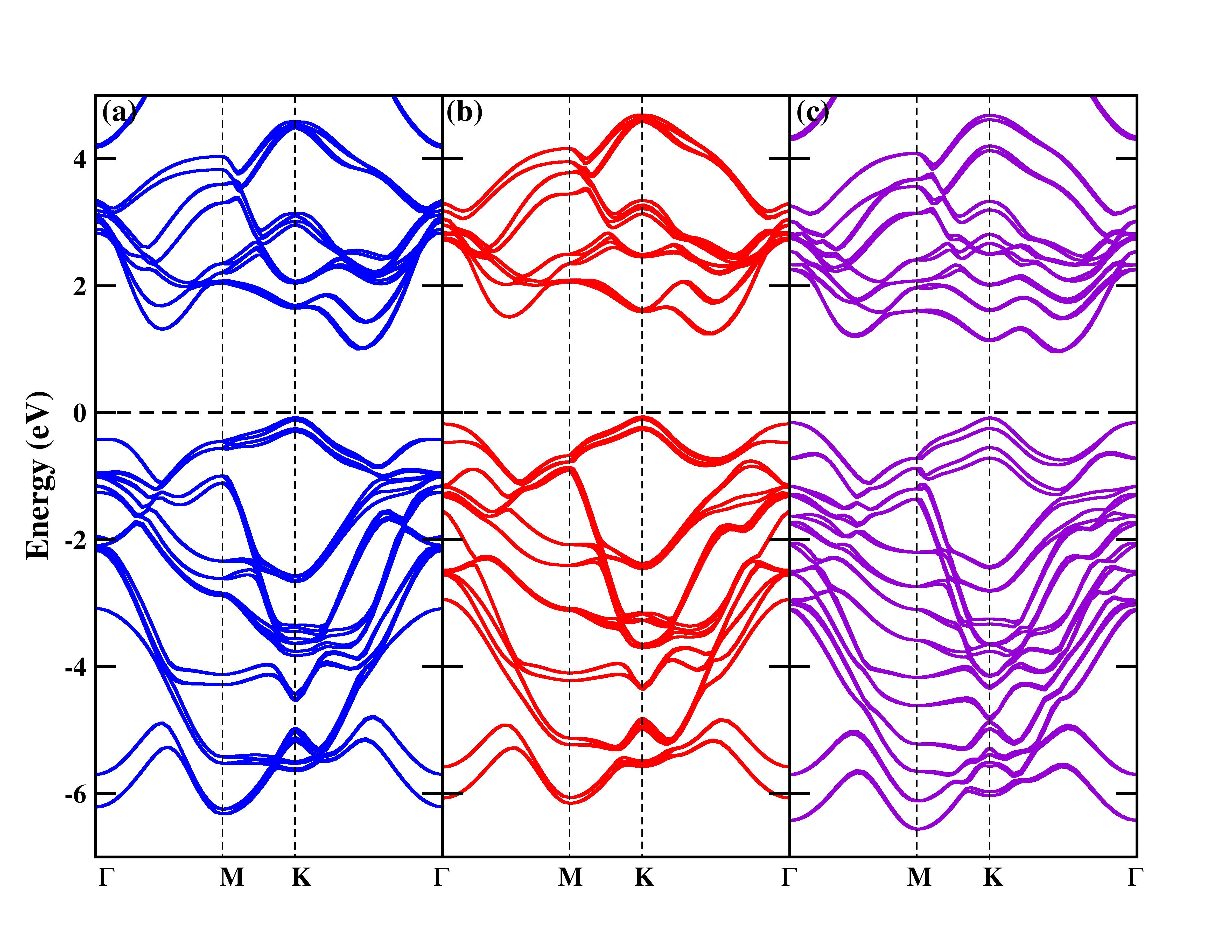}
\end{center}
\caption{ Band structure along the high symmetry $k$-points for the bilayers of (a) MoSe$_2$  (b) Janus I (c) Janus II. The Fermi energy is set to zero in the energy scale.}
\label{fig:Fig2}
\end{figure}

\begin{figure}
\begin{center}
\includegraphics [width= 12 cm] {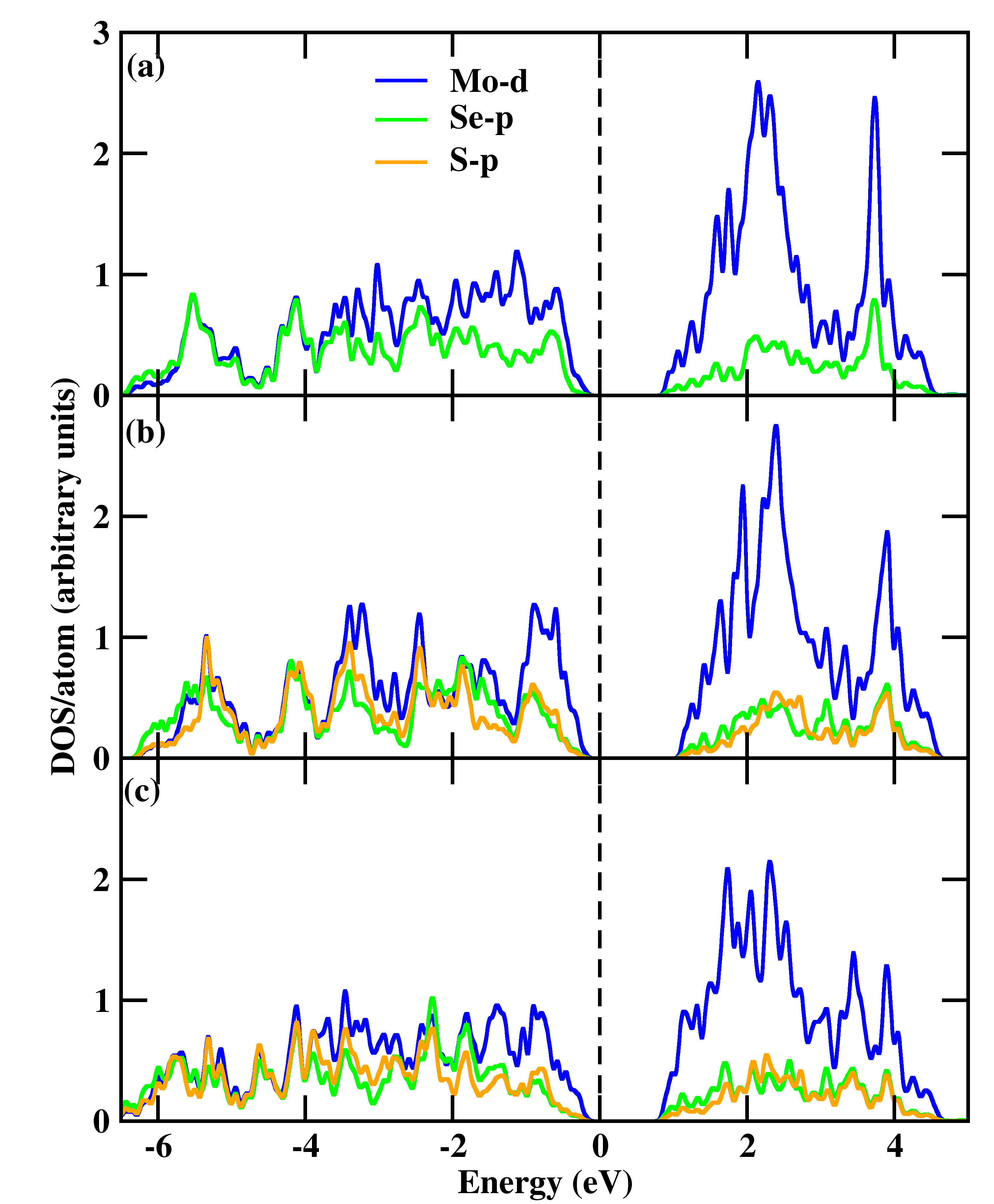}
\end{center}
\caption{The orbital projected denisty of states (DOS) is plotted for (a) MoSe$_2$ (b) Janus I and (c) Janus II. The Fermi energy is set to zero in energy scale.}
\label{fig:FigDOS}
\end{figure}

The electronic band structures of the sliding ferroelectric bilayers MoSe$_2$, Janus I, and Janus II along the high-symmetry path $\Gamma$–M–K–$\Gamma$, are presented in Fig. \ref{fig:Fig2} (a–c). All three systems exhibit an indirect bandgap, with variations in the positions of the valence band maximum (VBM) and conduction band minimum (CBM). From the orbital projected density of states (see Fig. \ref{fig:FigDOS}) we obtain that the VBM and CBM are predominantly composed of Mo-\textit{d} orbitals. The chalcogen \textit{p}-states (Se or S) are completely filled and lie well below the Fermi level, playing a minimal role in determining the low-energy electronic properties.
The VBM is located near the K point, while the CBM appears along the K-$\Gamma$ direction. The band gaps are of the order of 1 eV, 1.2 eV and 0.95 eV for MoSe$_2$, Janus I and Janus II respectively. In all three cases the values of the band gap are within the visible range, making these systems promising candidates for bulk photovoltaic applications.


\section{Wannier interpolation of the band structure}
\begin{figure}[H]
\begin{center}
\includegraphics [width= 15 cm] {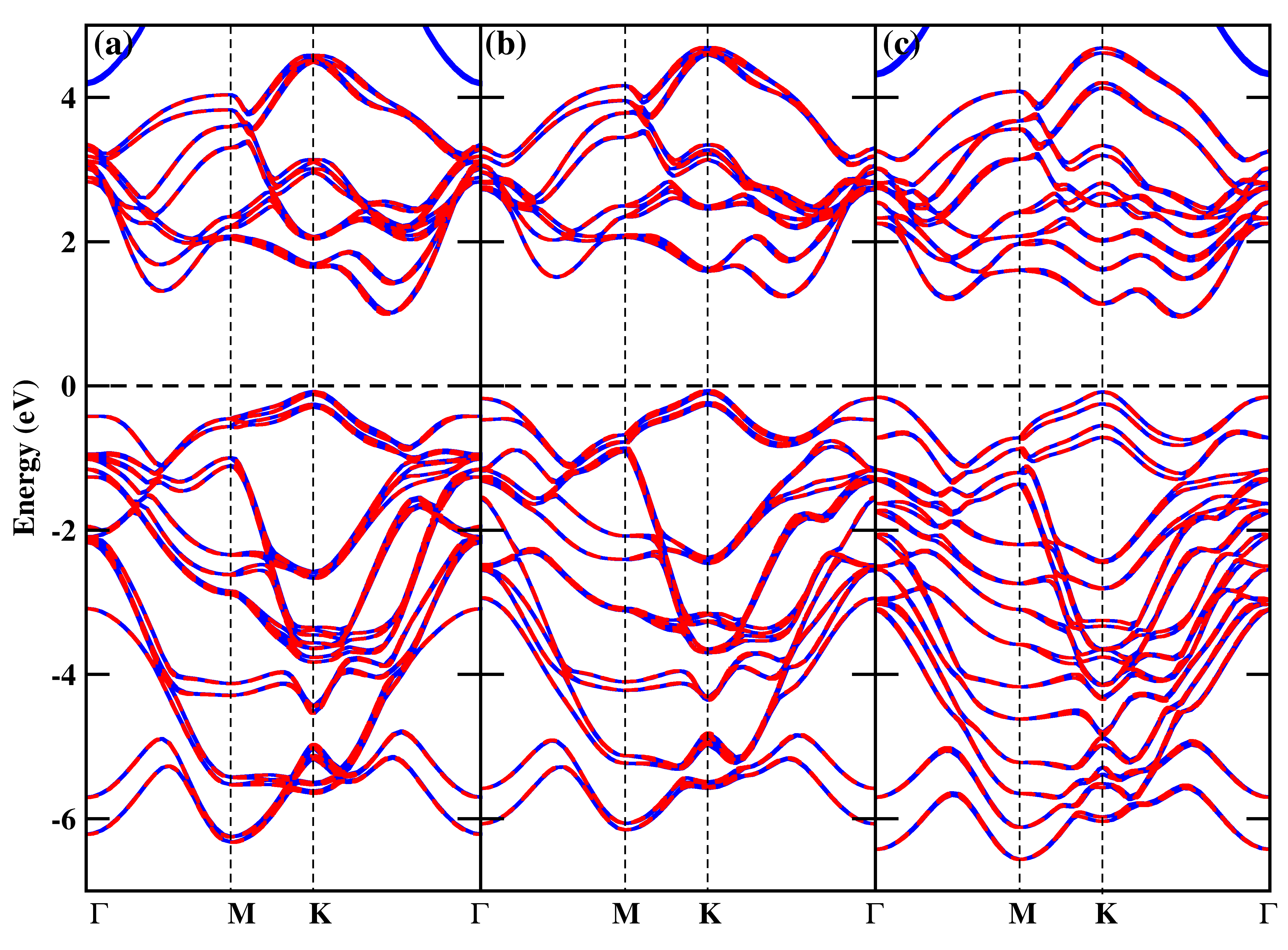}
\end{center}
\caption{The band structure plots along high-symmetry k-points for the bilayers of (a) MoSe2 (b) Janus I and (c) Janus II. The blue and red lines depict the DFT total band and the Wannierized bands, respectively. The Fermi energy is set to zero.}
\label{fig:FigW90}
\end{figure}

The DFT band structure of the bilayers, alongside the Wannier bands are displayed in Fig. \ref{fig:FigW90} . The DFT bands (blue lines) are well reproduced by the Wannier bands (red dotted lines) in a wide energy range around the Fermi level, namely from -6.5 to 4.5 eV.

\begin{figure}[H]
\begin{center}
\includegraphics [width= 15 cm, trim={5cm 0 4cm 0},clip] {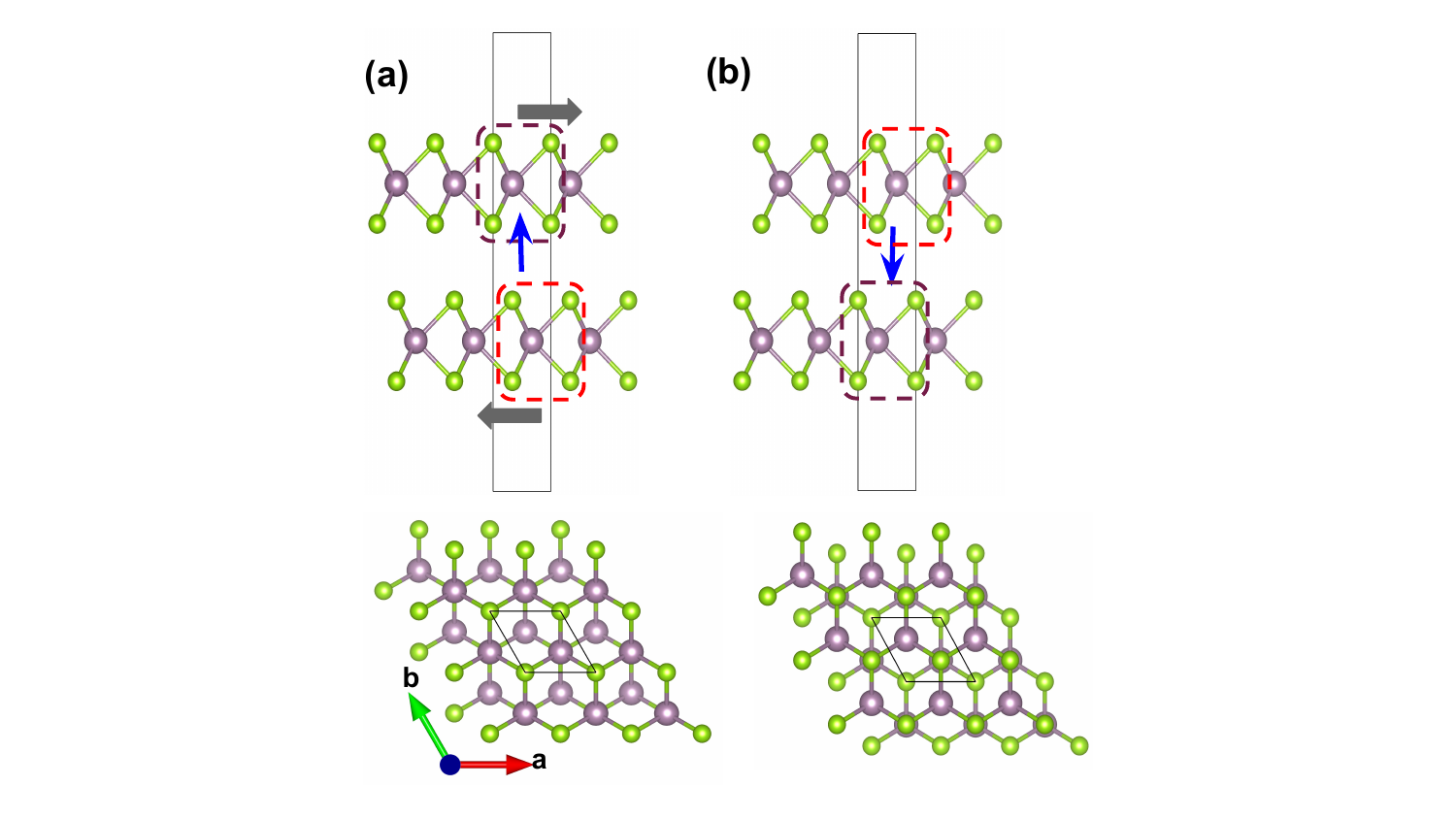}
\end{center}
\caption{Schematic representation of the sliding mechanism for MoSe$_2$ bilayer. The grey arrows represent the sliding direction of the monolayers. The blue arrows represent the direction of polarization. (a) and (b) are the structures with +P and -P polarization directions, respectively. The bottom panels represent the top view of the respective bilayers.}
\label{fig:Slide}
\end{figure}

\section{BPVE response for opposite polarization values}
\begin{figure}[H]
\begin{center}
\includegraphics [width= 12 cm] {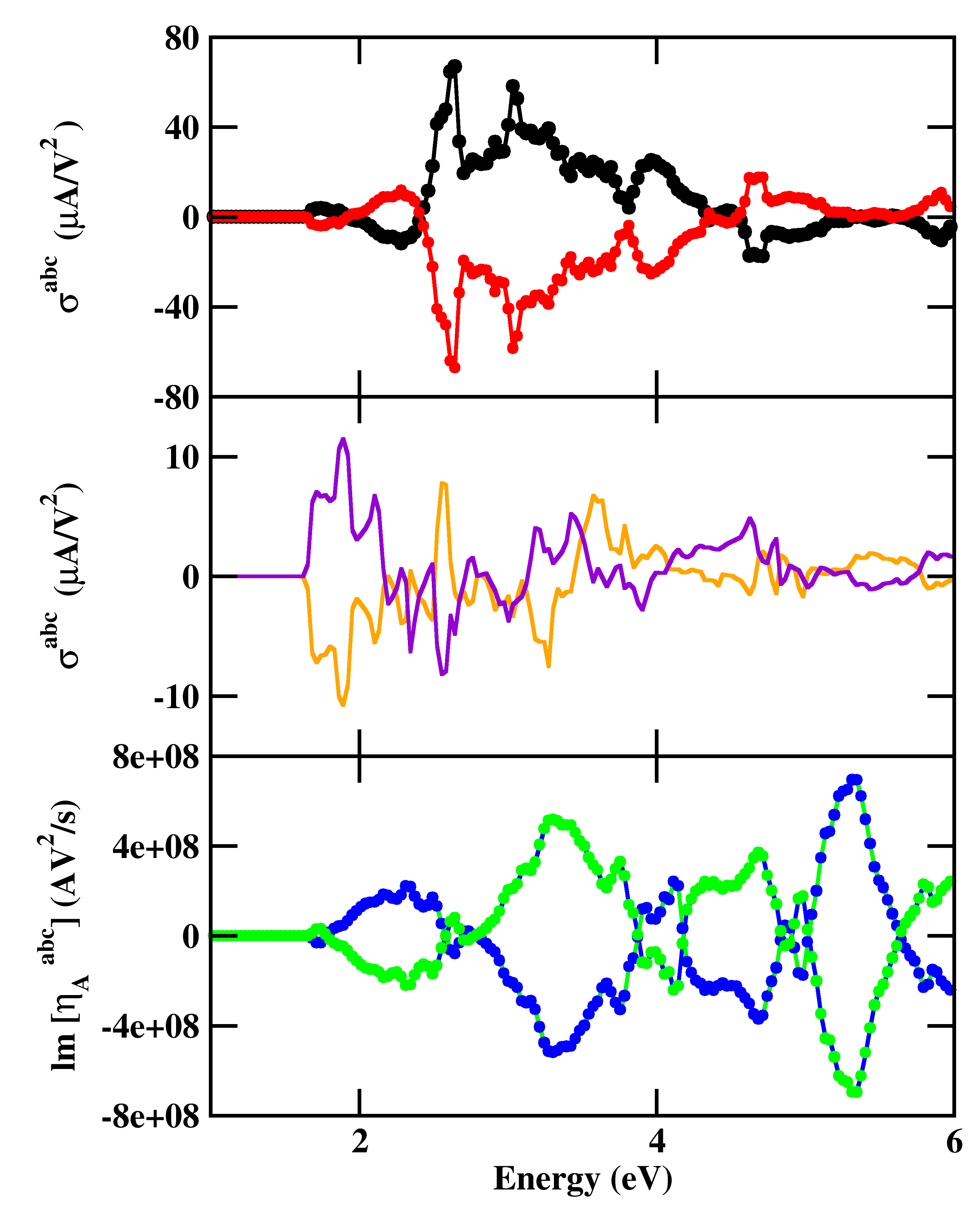}
\end{center}
\caption{The non-zero, independent components of the photoconductivity tensor is shown for (a) in-plane shift current components [$\sigma^{xxy}$(+P): black curve, $\sigma^{xxy}$(-P): black circles, $\sigma^{yyy}$(+P): red curve, $\sigma^{yyy}$(-P): red circles] (b) out-of-plane shift current components [$\sigma^{zxx}$(+P): violet curve, $\sigma^{zxx}$(-P): orange curve] (c) circular injection current components [$\eta^{xzx}$(+P): blue curve, $\eta^{xzx}$(-P): blue circles, $\eta^{yyz}$(+P): green curve, $\eta^{yyz}$(-P): green circles]. }
\label{fig:FigMs2}
\end{figure}

The mechanism for the sliding ferroelectricity is shown in Fig. \ref{fig:Slide}. As we slide the monolayers in opposite directions, we obtain structures with equal and opposite polarization values (+P and -P), as represented by the blue arrows. The evolution of the BPVE components, in bilayers with vacuum of $\approx$ 20 {\AA}, for opposite directions of the polarization, is shown in Fig. \ref{fig:FigMs2}. It can be seen that the out-of-plane components switch when the polarization is reversed, while the in-plane components do not.


\newpage
\section{ Structures for Janus II (slide) and Janus II'}
\begin{figure}[H]
\begin{center}
\includegraphics [width= 15 cm] {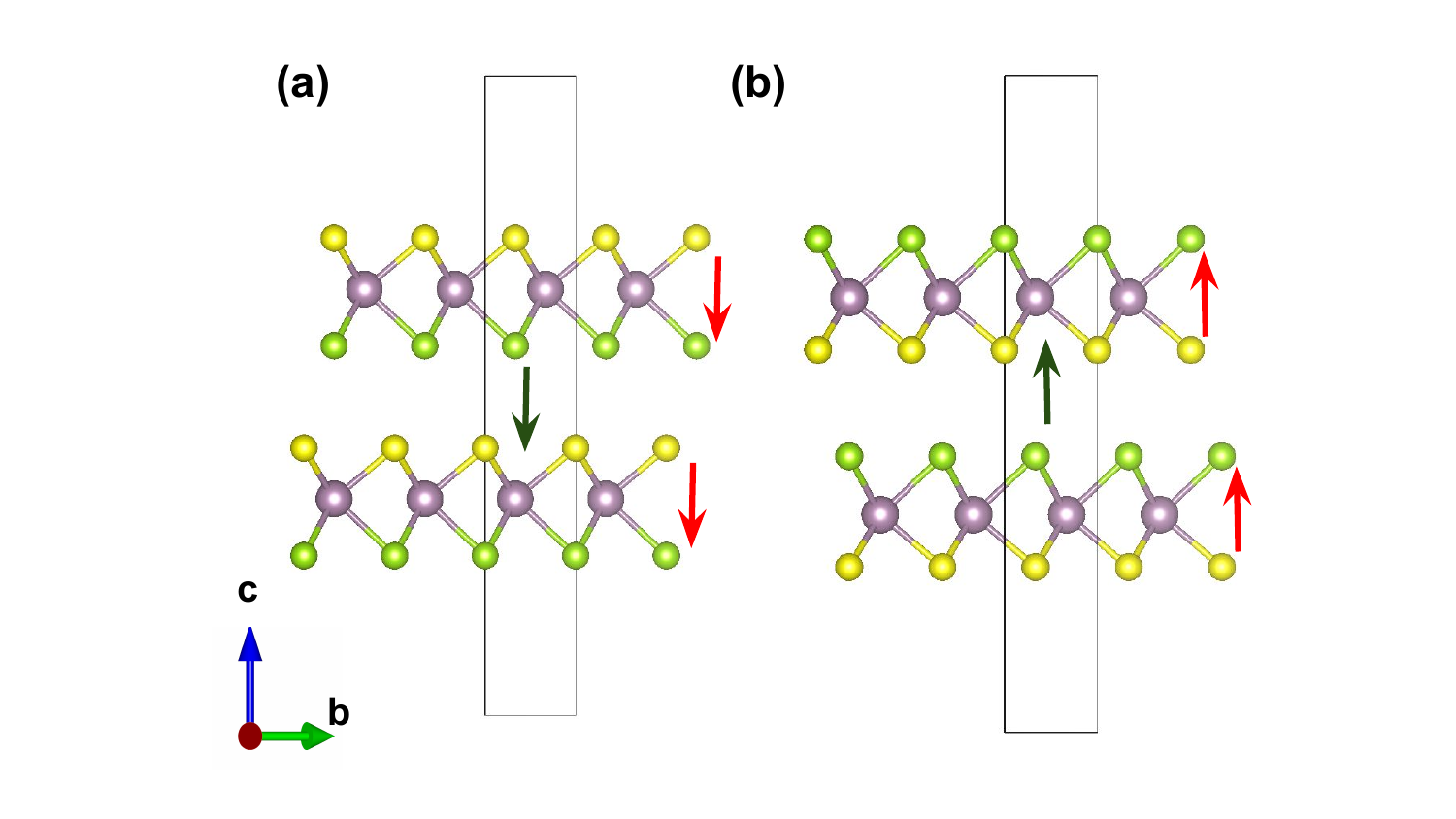}
\end{center}
\caption{The structure for (a) Janus II (slide) and Janus II'. The red and green arrows represent the directions for intralayer and interlayer polarization, respectively.}
\label{fig:FigJanus}
\end{figure}

Janus II (slide) represents the case where only the interlayer polarization is switched, induced by sliding the Janus II monolayers with respect to each other, as shown in Fig. \ref{fig:FigJanus} (a). Janus II' represents the case where only the intralayer polarization is switched, which is obtained by exchanging the S and Se atoms in the individual monolayers, as shown in Fig. \ref{fig:FigJanus} (b).

\end{document}